\documentclass[aip,apl,showpacs,reprint,amsmath,amssymb,superscriptaddress]{revtex4-1}

\usepackage{graphicx}
\usepackage{dcolumn}
\usepackage{bm}
\usepackage{array}
\usepackage{natbib}
\usepackage{epstopdf}
\usepackage{multirow}
\usepackage{color}
\usepackage{enumerate}

\begin{document}

\title{Long-term drift of Si-MOS quantum dots with intentional donor implants}

\author{M.~Rudolph}
\affiliation{Sandia National Laboratories, Albuquerque, NM, 87185, USA}
\author{B.~Sarabi}
\affiliation{National Institute of Standards and Technology, Gaithersburg, MD, 20899, USA}
\author{R.~Murray}
\affiliation{National Institute of Standards and Technology, Gaithersburg, MD, 20899, USA}
\author{M.S.~Carroll}
\affiliation{Sandia National Laboratories, Albuquerque, NM, 87185, USA}
\author{Neil~M.~Zimmerman}
\affiliation{National Institute of Standards and Technology, Gaithersburg, MD, 20899, USA}

\date{\today}

\begin{abstract}

Charge noise can be detrimental to the operation of quantum dot (QD) based semiconductor qubits.  We study the low-frequency charge noise by charge offset drift measurements for Si-MOS devices with intentionally implanted donors near the QDs.  We show that the MOS system exhibits non-equilibrium drift characteristics in the form of transients and discrete jumps that are not dependent on the properties of the donor implants.  The equilibrium charge noise indicates a $1/f$ noise dependence, and a noise strength as low as $1~\mathrm{\mu eV/\sqrt{Hz}}$, comparable to that reported in more model GaAs and Si/SiGe systems (which have also not been implanted).  We demonstrate that implanted qubits, therefore, can be fabricated without detrimental effects on long-term drift or $1/f$ noise.

\end{abstract}

\maketitle

For semiconductor quantum dot (QD) based qubits, charge noise has been identified as a critical parameter influencing the fidelity of the qubit \cite{coish2005}.  Measurements indicate that the noise is $1/f$-like, which results in a quasi-static dephasing signature, where the low-frequency noise components (with respect to the qubit operations) dominate the dephasing \cite{dial2013, eng2015}.  We explore the impact of implanting donors near the active QD regime of Si-SiO$_2$ devices on the strength of the low frequency charge noise.  In particular, we measure the charge offset drift $Q_0$, which has its biggest impact on the possible integration of devices \cite{zimmerman2008}.  Previous studies have shown an order of magnitude reduction in $Q_0$ for a Si-SiO$_2$ MOS system compared to dielectric stacks that include AlO$_\mathrm{x}$ (present with Al top gates) \cite{zimmerman2008, zimmerman2014}.  However, the influence of donors in QD systems has not been studied.  The process of donor implants is necessary for donor-based qubits, which have recently demonstrated extremely long coherence times \cite{muhonen2014}.  The QD qubit motivating this work is a QD-donor double well system operated in a two-electron singlet triplet scheme \cite{harvey-collard2017, rudolph2017}.  This hybrid system has the potential to provide very long coherence times, as have been shown in single donor qubits.  In addition, searching for the elusive donor-donor coupling \cite{gorman2016} is bypassed by the tunability of the QD-donor coupling.  Due to the statistical straggle of donor implantation, the exact location of a donor can be estimated to within only tens of nanometers, so more donors than are necessary are implanted to ensure that a donor is placed in a target zone \cite{bielejec2010, singh2016}.  We study the effects of various quantities of implanted donors on the low frequency charge noise and show that, for certain devices, the process of implanting donors does not affect the charge noise.

\begin{figure}
\includegraphics[width=86mm]{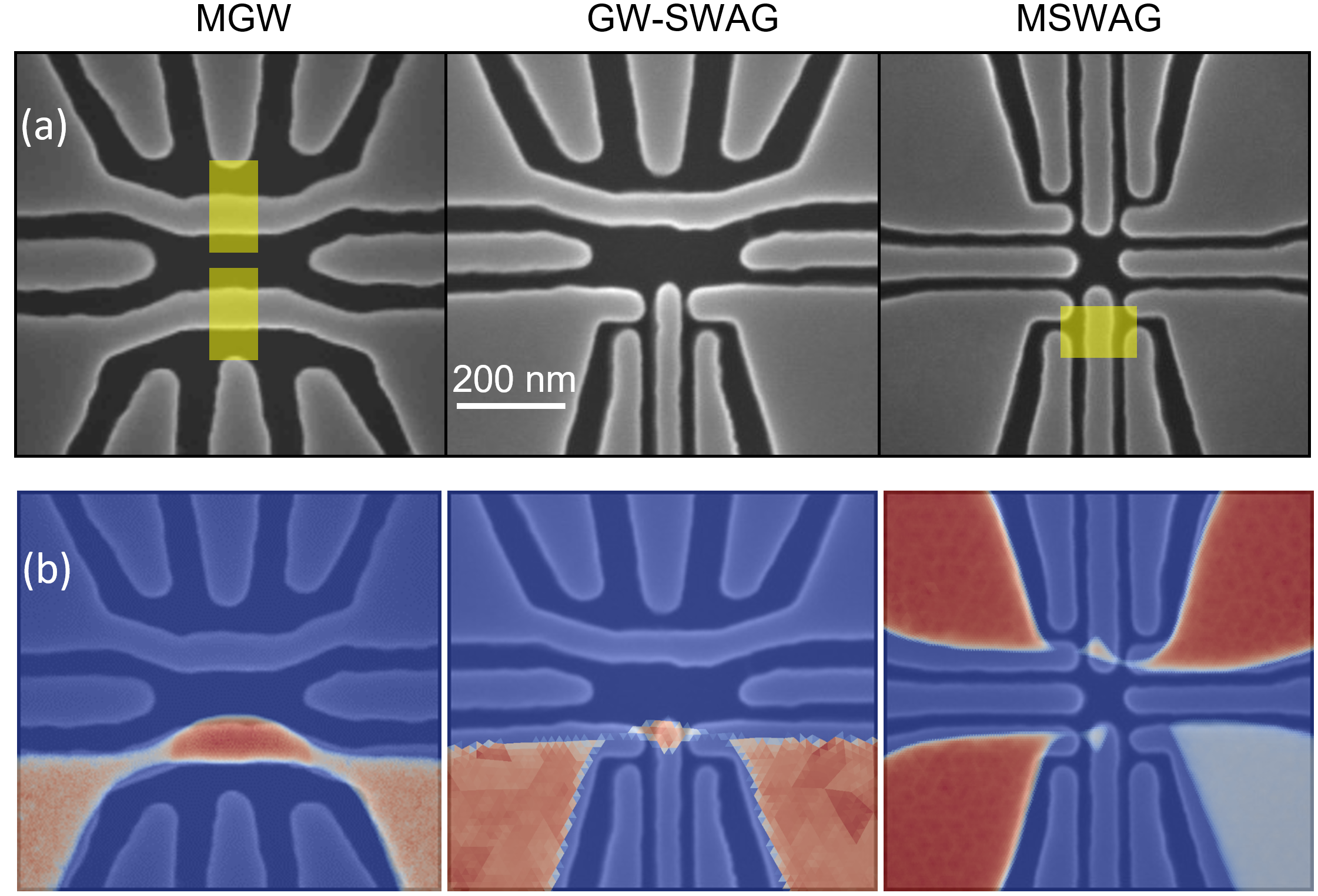}
\caption{(a) Scanning electron micrographs of the three device designs measured.  The light gray are the poly-Si gates, and the yellow regions are where donors are implanted.  (b) Simulated electron density (red regions) for each of the devices.  The lower half of the device is turned on for MGW and GW-SWAG, while for MSWAG both the lower QD and upper charge sensor are on.}
\label{fig:sem}
\end{figure}

Long-term drift was measured for six devices.  The silicon source material and devices were all fabricated at Sandia National Labs, with minimal differences in the process flow \cite{tracy2009} (fabrication details in Appendix A).  Two of the devices have isotopically enriched $\mathrm{^{28}Si}$ substrates, while the other four are naturally abundant silicon.  The material stack consists of the Si substrate, a $35~\mathrm{nm}$ SiO$_2$ gate oxide, and a $200~\mathrm{nm}$ poly-Si gate.  The poly-Si layer is patterned with one of the geometries shown in Figure~\ref{fig:sem}(a), where each device is capable of forming two QDs, one in the lower half of the device and one in the upper half.  Simulations of the electron density for actual operating voltages are shown in Fig.~\ref{fig:sem}(b) to highlight the differences in QD sizes and proximity to the implanted donors.  Next, some devices received donor implants confined near the QD regions by a pattern mask (yellow boxes).  Both $120~\mathrm{keV}$ Sb and $45~\mathrm{keV}$ P ions were studied, which have similar target depths of $28~\mathrm{nm}$, but the Sb implants have a narrower straggle ($18~\mathrm{nm}$ compared to $25~\mathrm{nm}$).  The total number of donors implanted differs between the devices due to a difference in both the implant fluence and the implant window size.  In addition, the location of the implant differs between devices, with some devices having the implants around the QD being studied, while other devices have implants on the other half of the device around the inactive QD.  Table~\ref{tab:all} provides the parameters for each of the devices studied.  For the number of implanted donors, we distinguish between the number of donors implanted near the active QD from donors implanted near the inactive QD (in parentheses).  All implanted devices received an activation anneal, while all devices received an identical post-fabrication forming gas anneal.

\begin{table*}
\begin{tabular}{c | c c c c c c}
Device & DA & DB & DC$\mathrm{^{a}}$ & DD\footnote[1]{$\mathrm{^{28}Si}$ device} & DE & DF \\
 \hline\hline
Geometry & GWSWAG & GWSWAG & MSWAG & MGW & MGW & MGW \\ 
Implant donor & - & - & P & P & Sb & Sb \\
Implant energy (keV) & - & - & 45 & 45 & 120 & 120 \\
Implant Fluence($10^{11}\mathrm{cm^{-2}})$ & - & - & 4 & 8 & 4 & 4 \\
Anneal ($\mathrm{^{\circ}C}$) & - & - & 1000 & 900 & 900 & 900 \\
Number of donors & 0 & 0 & 27 (13) & 60 (60) & 45 (0) & 0 (45) \\
Substrate Resistivity ($\mathrm{\Omega cm}$) & $>10^4$ & $>10^4$ & $>10^4$\footnote[2]{Resistivity of natural silicon substrate.  The resistivity of the $\mathrm{^{28}Si}$ epi-layer is not measured} & $>10^4$$\mathrm{^b}$  & $>10^4$ & $>10^4$  \\
$T (K)$ & 4.1 & 4.1 & 0.2 & 0.2 & 2.3 & 2.3 \\
$\sigma_0~(\mathrm{\mu eV})$ & $40\pm15$ & 70 & 4 & - & $9\pm3$ & $12\pm4$ \\
$e\sigma_0/E_C~(10^{-3}e)$ & 15 & 15 & 0.5 & - & 9 & 12 \\
$\tau_{tr}~\mathrm{(hr)}$ & 6 & 11, 3.3 & - & - & 17 & $>30$ \\
$\beta$ & $2.65\pm0.24$ & $1.29\pm0.12$ & $1.07\pm0.13$ & - & $0.85\pm.015$ & $0.89\pm0.18$\\
$\sqrt{\sigma_\beta}~(\mu eV)$ & - & $7\pm3$ & $1\pm0.6$ & - & $7\pm4$ & $5\pm3$ \\
\hline\hline
\end{tabular}
\caption{Compilation of device parameters.  The number of implanted donors is separated into donors near the measured SET and, in parentheses, donors on the opposite device of the SET.  Measurements performed at the indicated temperature exhibited a standard deviation in the chemical potential drift and charge offset drift of $\sigma_0$ and $e\sigma_0/E_C$, respectively.  For devices where the lever arm was not measured, bounds for $\sigma_0$ are estimated from values of the charging energy measured for other devices with identical geometries.  For instances where a transient drift can be fit to an exponential, the decay paramter $\tau_{tr}$ is reported.  The exponent of the spectral density $\beta$ and the noise strength $\sqrt{\sigma_\beta}$ at $1~\mathrm{Hz}$ are provided.}
\label{tab:all}
\end{table*}

\begin{figure}
\includegraphics[width=82mm]{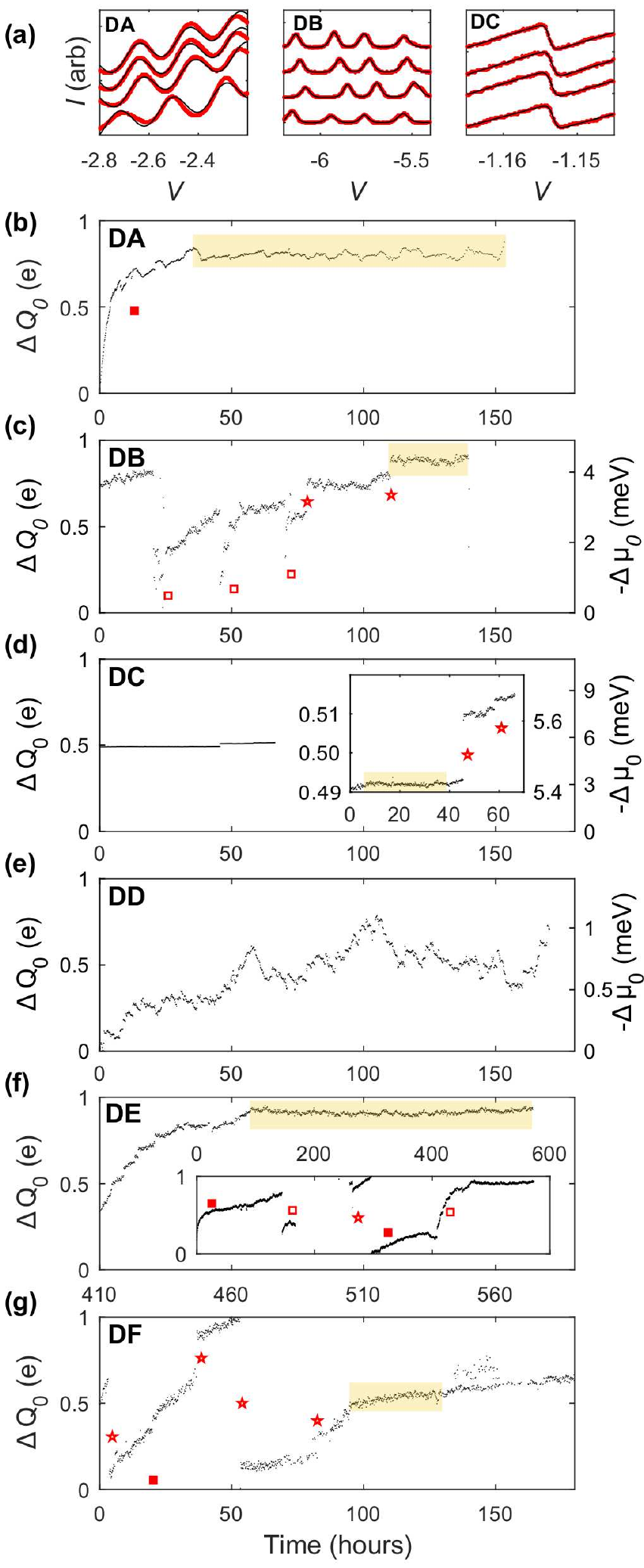}
\caption{(a) Example fits to the chemical potential position for electron transport measurements (left, Eq.~\ref{eq:sin}; middle, Eq.~\ref{eq:cosh}), and for charge sensed measurements (right, Eq.~\ref{eq:dicarlo}).  Traces are taken a day apart, and horizontal shifts represent $\Delta V(t)$.  (b-g) Charge offset drift for devices DA-DF. Occurrences of transient relaxation due to thermal shock ($\textcolor{red}{\blacksquare}$) and external shocks to the measurement apparatus ($\textcolor{red}{\square}$) as well as discrete charge redistributions ($\textcolor{red}{\bigstar}$) are annotated.  Regions where local fluctuations of charges dominate are highlighted yellow, and are where $\sigma_0$ is measured.}
\label{fig:drift}
\end{figure}

To measure the long-term drift characteristics of our devices, a QD is tuned up in the lower half of the device.  For Devices DC and DD, the upper QD is also present, but the drift is only measured on the lower QD.  The simulated electron densities of the devices during operation are displayed in Fig.~\ref{fig:sem}(b).  The relative position of the QD chemical potential at fixed charge occupation is measured either through the transport Coulomb blockade signature in the lower QD, or it is measured by the charge sensed response of the upper QD.  The functional form of the transport measurement depends on the ratio $E_C/k_{B}T$.  For values less than 5, the Coulomb blockade is not robust, and the data is approximated by a sine function \cite{zimmerman2008}, where 
\begin{equation}
I(V,t)=I_0 + I_1V + A\sin[2\pi (C/e)(V+\Delta V(t))] .
\label{eq:sin}
\end{equation}
Here, $V$ is the gate voltage applied, and $C$ is the capacitance of the gate to the QD.  In this region, the conductance has contributions from the quantum Coulomb blockade effect and the classical transistor turn on, the latter of which is approximated by a linear voltage response to the small voltage swings applied.  The chemical potential position is assigned to the phase $\Delta V(t)$ of the sine fit.  For values of $E_C/k_{B}T\gtrsim5$, where $E_C$ is the QD charging energy, the Coulomb blockade is robust and the conductance goes to zero.  The data is fit to \cite{meirav1990}
\begin{equation}
I(V,t)=A\cosh^{-2}[B(V+\Delta V(t))] .
\label{eq:cosh}
\end{equation}
The chemical potential is defined by the peak center.  For the charge sensing measurement, the chemical potential is extracted by a fit to the center of a Fermi-Dirac distribution \cite{dicarlo2004}
\begin{equation}
I(V,t)=I_0 + I_1V + \frac{A}{1+\exp[B(V+\Delta V(t))]} .
\label{eq:dicarlo}
\end{equation}
A linear background approximating the direct charge sensor response to the gate is included.  Examples of all three types of data are presented in Fig.~\ref{fig:drift}(a).  The measurement is repeated approximately every 10 minutes for multiple days to track changes in the position of the QD chemical potential.  The traces in Fig.~\ref{fig:drift}(a) are offset for clarity with each scan separated by a day, with later scans on the top.

The extracted voltage shifts $\Delta V(t)$ are device and geometry specific, and are converted to charge offset drift. For many-electron QDs, the regular period of the Coulomb blockade peaks indicates the voltage required to change the QD occupation by one electron, and so the QD chemical potential position can be expressed in units of charge, where $\Delta Q_0 (t) = e \Delta V_N(t) / (V_{N+1}-V_{N})$, with $N$ the QD occupation \cite{zimmerman2008}.  Figures~\ref{fig:drift}(b-g) show the long-term charge offset drift for each of the devices.  It is also useful to convert the charge offset drift to a chemical potential drift, which is important for relating the noise to qubit performance.  If the gate lever arm $\alpha$ is known, the chemical potential drift is simply $\Delta \mu_0 (t) = -\alpha \Delta V_N(t)$, where $\alpha$ has units of eV/V.  The two can be related by $\Delta \mu_0 = -(E_C/e) \Delta Q_0$.  For instances where $\alpha$ is accurately known, the chemical potential drift is indicated on the right axes (note negative sign).

In the context of qubit dephasing, the charge noise is best denoted in units of energy, which directly describes the fluctuations of the QD energy levels induced by nearby charge noise.  Experiments measuring the charge noise by qubit dephasing typically observe quasi-static values from 1-$10~\mathrm{\mu eV}$ \cite{dial2013, eng2015}.  In addition to qubit gate fidelity, low-frequency charge noise can also affect the qubit readout fidelity.  To read out the qubit state, many schemes utilize a single electron transistor (SET) to detect the charge state of the QD, which can infer the qubit state \cite{dicarlo2004}.  Low-frequency charge noise can modify the readout sensitivity and require frequent calibrations of the read-out circuit.  The maximum sensitivity of the SET is often constrained by the limiting energy scale in SET (i.e. the electron temperature or bias voltage), and is typically $<100~\mathrm{\mu eV}$.  This scale sets a maximum amount of low frequency drift that can be tolerated before the read-out circuit needs to be retuned.  For this application, describing the charge noise in units of energy allows direct comparison to the energy scales of the read-out circuit.  The same is true for devices that use a quantum point contact for the read-out electrometer \cite{elzerman2003}.

On the other hand, the low-frequency charge noise has been described by the charge offset drift parameter, which extracts the capacitively-coupled displacement charge induced on the SET by the environmental charge noise \cite{zimmerman2008, zimmerman2014}.  This quantity is appropriate when the read-out SET is tuned such that the Coulomb blockade is quasi-sinusoidal, which happens when $k_BT\sim E_C/5$ and when the SET tunnel barriers are tuned to be open.  Here $E_C$ is the SET charging energy.  Previous measurements operated in this regime.  The benefit of reporting the charge offset drift over the energy drift in a QD device is that it may provide a better figure of merit in comparing the noise between different devices or materials.  The charge offset drift is defined as $Q_0(t) = -e/C_\Sigma \sum_i C_{m,i}(t)$, where $C_{m,i}(t)$ is the time dependent mutual capacitance of a fluctuating charge $i$ and $C_\Sigma=e^2/E_C$ is the total capacitance of the QD.  The origin of the fluctuating charges $i$ here is not important, and may include charges in the gate, in the oxide, or in the substrate.  The energy drift is defined as $\mu_0(t) = -Q_0(t)E_C/e = (e/C_\Sigma)^2 \sum_i C_{m,i}(t)$.  To first order, both $C_m$ and $C_\Sigma$ scale with the QD size, which makes $Q_0(t)$ insensitive to changes in the geometry of the QD.  This suggests that $Q_0(t)$ is a better quantity to compare the external charge noise between different devices.

To summarize, we suggest that expressing the charge offset drift in energy units is particularly useful in the contexts of i) dephasing in qubits and/or ii) read-out with a SET or point contact that has a cusp-like response, where the width of the cusp is set by an energy scale.  In addition, units of energy will provide a natural comparison between charge noise and other mechanisms that can coupled to the qubit system, such as phonons \cite{beaudoin2015}.  We suggest that expressing the drift in charge units is useful in discussing i) digital integration of multiple qubits and ii) comparing the behavior of different devices, especially from different groups.

The characteristics of the drift observed (Fig.~\ref{fig:drift}) in the devices can be separated into three categories.
\begin{enumerate}
\item A transient relaxation.  This can occur after a thermal shock such as a cooldown.  Measurements were performed within a few hours of cooldown for devices DA, DE, and DF, and the chemical potential position initially follows a quasi-exponential relaxation until the saturation after about 2 days.  Regions of thermally induced transients are indicated by ($\textcolor{red}{\blacksquare}$).  Transient relaxation also occurs after a non-equilibrium charge reconfiguration due to an external voltage shock.  This is evident in device DB where the spikes observed at hours 24, 48, and 72 were induced by work being performed near the cryostat.  Device DE has two shock events as well.  The relaxation here is also on the order of a day.  Electrostatically induced transients are indicated by ($\textcolor{red}{\square}$).
\item Isolated discrete jumps.  Reconfiguration of isolated charges in the device can induce a chemical potential shift with no subsequent relaxation.  Two such events are visible in device DB at hours 75 and 110.  Devices DC, DE, and DR also display this behavior, which are indicated by ($\textcolor{red}{\bigstar}$).  
\item Local fluctuations about a stable mean value are present in all measurements and originate from the charge noise induced by remote charge reconfigurations and gate noise.  Examples of these are highlighted in the figure.  The size and spectrum of these local fluctuations measure the intrinsic equilibrium charge noise of the system.  For device DA, there exists a dominant slow two-level-fluctuator, while the equilibrium fluctuations for the other devices are on faster time scales.
\end{enumerate}
Device DD shows significantly different behavior than the others, with a slow, non-monotonic drift dominating the charge offset measurement.  This device will be excluded in the following discussion and revisited at the end. 

Before discussing these results in detail, we wish to put the charge offset drift in the context of previous measurements.  As discussed above, the charge offset drift is an important feature for the prospect of integration of single electron devices in general, and has been measured in a variety of devices \cite{zimmerman2008, stewart2016}.  In particular, the general observation has been that phenomena such as the transient relaxation and slow, non-monotonic drift have not been observed in all-Si devices fabricated from silicon on insulator (SOI) substrates, which have demonstrated background levels of drift of $e\sigma_0/E_c \lesssim 0.03~\mathrm{e}$.  Here $\sigma_0$ is the standard deviation of $\Delta \mu_0(t)$.  We note that infrequent discrete shifts have been observed.  Si devices with aluminum gates have shown non-monotonic drift \cite{zimmerman2014}, with $e\sigma_0/E_c \approx 0.15~\mathrm{e}$; while devices based on $\mathrm{Al/AlO_x/Al}$ tunnel junctions have large instabilities with non-monotonic drift, transient relaxation, and frequent discrete shifts, with overall fluctuations $e\sigma_0/E_c >1~\mathrm{e}$.  These observations have been interpreted as suggesting that the general behavior and magnitude of the charge offset drift depends on the quality of the insulators surrounding the quantum dot.

However, our observations in all-Si devices with bulk Si substrates, as shown in Fig.~\ref{fig:drift}, of transient relaxation and (in one case) slow, non-monotonic behavior, show different behavior than those previous results in all-Si SOI-based devices.  Both all-Si architectures exhibit infrequent discrete shifts, which are often correlated with external impulses on the measurement apparatus.  It is apparent from this difference that the previous general conclusion that the charge offset drift depends only on the materials quality of the nearby insulators is insufficient, and that the device architecture may also play a role.  We now proceed to discuss the transient relaxation, discrete shifts, and local fluctuations in our devices.  The results are summarized in Table~\ref{tab:all}.

\begin{figure}
	\includegraphics[width=86mm]{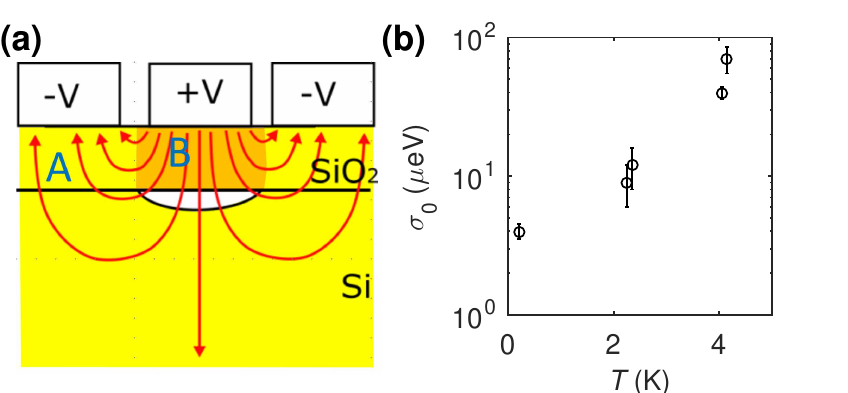}
	\caption{(a) Qualitative electric field profile near the QD. In region A (yellow) the electric fields point away from the QD, which in region B (orange) the electric fields point toward the QD. (b) The standard deviation of the local chemical potential fluctuations as a function of electron temperature.}
	\label{fig:efield}
\end{figure}

\paragraph*{Transient Relaxation:}
For the transient relaxation present, the direction of the relaxation is the same in all devices.  The transient decay time $\tau_{tr}$ for the relaxation, computed by an exponential fit $e^{-t/\tau_{tr}}$, is found to be between 3 and 17 hours for the 3 instances.  Both devices DE and DF have an instance of a longer relaxation of $>30$ hours, but do not reliably fit an exponential form.  All devices were measured in different cryostats and have different donor implant parameters, suggesting an intrinsic origin of the transient behavior.  Notably, devices DA and DB had no donor implants, while device DE was implanted, indicating that the addition of donors near the QD does not affect the transient behavior.  In all cases, the relaxation lowers the chemical potential of the QD.  This can occur when negative (positive) charges migrate away from (toward) the QD, requiring the electric field point towards the QD (region B in Fig.~\ref{fig:efield}(a)).  In our lithographic devices, the confining lateral electric fields point away from the QD at the interface (region A in Fig.~\ref{fig:efield}(a)), ruling out the reorientation of slow interface traps.  We offer two hypotheses for the transient relaxation.  
\begin{itemize}
	\item In the SiO$_2$, the fields closest to the QD point towards it while farther away they can point away from the QD (Fig.~\ref{fig:efield}(a)).  The capacitive effect of charge motion would be dominated by the closer regions, consistent with lowering the chemical potential, and indicating that the charge motion inducing the transient relaxation can occur in the SiO$_2$ (region B).  Because presence of the donors in the Si does not significantly affect the relaxation, this would suggest that any damage induced in the SiO$_2$ from the implant process (an anneal step is performed to minimize damage) does not affect the relaxation.
	\item When the device is first turned on, a depletion region must be formed in the substrate before electrons can accumulate at the interface.  With our low-doped substrate, the depletion region is about $1~\mathrm{\mu m}$ at equilibrium.  In order to grow the depletion region, electrons must be injected into the substrate to neutralize the acceptor atoms.  At low temperatures, the substrate is frozen out (insulating), so it is possible that the rate at which the depletion region can respond to large voltage swings is limited due to electrons tunneling through the substrate from the implanted ohmic regions \cite{simoen1993a, simoen1993b}.  This mechanism would induce transients lowering the chemical potential for sudden positive voltage changes, which is consistent with normal operations during device turn on after cooling down.  This is independent of any deliberate donor implantation performed, and would exist in all devices.  The magnitude of the effect may be dependent on the amount of area enclosed by positive voltages (size of QD and reservoir gates).  In addition, SOI-based devices may suppress this effect as the depletion region is significantly reduced compared to bulk-Si based devices, so there are fewer background dopants to equilibrate.
\end{itemize}

\paragraph*{Discrete Shifts:}
The large discrete shifts observed in the QD chemical potential (device DB at hours 75 and 110, device DC at hours 45 and 58, device DE at hour 280, device DF at hours 8, 40, 52, and 80) are also all a lowering of the QD chemical potential.  Note that discrete shifts are observed in both implanted and non-implanted devices.  If an electron tunnels between two isolated interface trap sites, the observed shifts would be caused by an electron tunneling away from the QD.  However, the electric field would detune two trap sites such that there would be a larger probability of electrons tunneling toward the QD (assuming a random distribution of trap levels in both space and energy), contrary to our observations.  The discrete jumps in our devices are more likely to be caused by isolated interface electrons tunneling to the leads either directly or via cotunneling through the QD, leaving behind an ionized hole trap in the SiO$_2$.  One would expect these discrete transitions to saturate, similar to the transient behavior, after some time scale.  However, the individual jumps present are separated by $>10$ hours, and there are not enough discrete transitions to gather statistics and define a time scale for saturation.  As with the transient behavior, the discrete features are observed in both implanted and non-implanted devices, suggesting a mechanism intrinsic to the MOS system and not the implant parameters.

\begin{figure*}
	\includegraphics[width=180mm]{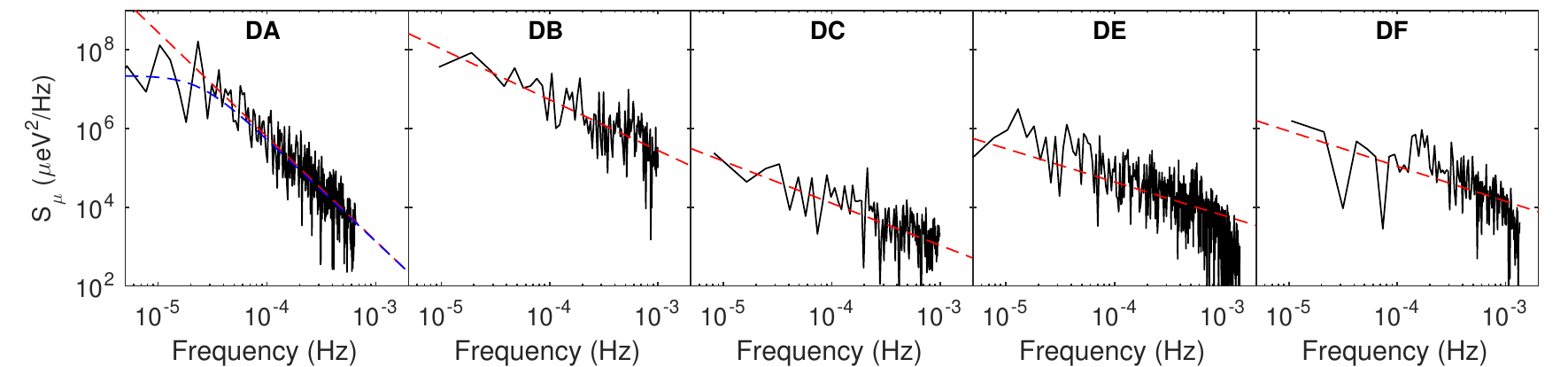}
	\caption{Power spectral density of the chemical potential drift for all devices, excluding DD.  The dotted lines are power law fits, which are tabulated in Table~\ref{tab:all}.}
	\label{fig:psd}
\end{figure*}

\paragraph*{Local Fluctuations:}
In between the transient and discrete shifts in the chemical potential, the chemical potential drift is stable about a mean value.  The fluctuations about this mean are presumably caused by metastable charge fluctuations in the device.  We compute both the standard deviation $\sigma_0$ and the spectral density of these local fluctuations.  The data used in these calculations are indicated by the high-lighted regions of Fig.~\ref{fig:drift}(b-g).  Since there may still be some slow transient behavior in these regions, a quadratic background is removed before the analysis.  The spectral density is computed by the Fourier transform of the fluctuations, which are shown in Fig.~\ref{fig:psd}.  Power law fits of the form $S_\mu = \sigma_\beta[1~\mathrm{Hz}/f]^\beta$ are applied to each spectrum (extracted parameters are in Table~\ref{tab:all}) and reveal that all devices (both implanted and non-implanted) except DA exhibit a $1/f$ noise dependence (device DB has $\beta$ slightly higher than 1, which may be due to a quadratic background not being sufficient to describe the residual slow drift). In MOS systems, one typically observes $1/f$ noise, whose origin can be attributed to an ensemble of a random distribution of two-level fluctuators \cite{machlup1954}.  Since fluctuators are temperature dependent \cite{beaudoin2015}, this model would predict an increase in the noise strength as temperature increases, which has been observed in GaAs and Si-based QD systems \cite{dial2013, freeman2016}.  In Fig.~\ref{fig:efield}(b), the variance of the chemical potential drift $\sigma_0$ for each device is plotted against their electron temperatures.  While direct comparison of the noise strength between different devices introduces unknown errors due to differences in specific disorder configurations in each device, a monotonic increase in noise is observed over the sample set of five devices.  The standard deviation in the charge offset drift $e\sigma_0/E_C$ and the spectral noise strength $\sigma_\beta$ also exhibit a monotonic increase with temperature.  We note that shot noise, Johnson noise, and instrument noise for these measurements are estimated to combine for, at most, a contribution of $0.04~\mathrm{\mu eV^2/Hz}$, which is well below the levels measured.

To improve the performance of a QD qubit system, it is important to minimize the charge noise.  The charge noise measured in QD qubit systems is often of $1/f$ nature, and the best reported values of the noise strength $\sqrt{\sigma_1}$ (where $\sigma_1=\sigma_\beta$ for $\beta=1$) have been in the range of 1-$10~\mathrm{\mu eV}$ at $1~\mathrm{Hz}$ at temperatures of $\sim0.1~\mathrm{K}$ for unimplanted GaAs, Si/SiGe, and Si-MOS devices \cite{dial2013, eng2015, freeman2016}.  At similar temperatures, we measure for device DC $\sqrt{\sigma_1}=1~\mathrm{\mu eV}$, which lines up very well with this strength.  The process of intentionally introducing donors near the QD does not significantly degrade the charge noise characteristics, and in fact the noise strength is similar to non-implanted MOS devices as well as nominally cleaner SiGe and GaAs devices \cite{dial2013, eng2015}.

For device DA, the increased low frequency noise is due to a dominant two-level-fluctuator present. Since the QDs are zero-dimensional objects and only sample a relatively small spatial distrubtion of fluctuators, one can expect that there is a reasonable chance for a dominant fluctuator to exist that will skew the random distribution of fluctuators that produces the $1/f$ noise.  The noise spectrum of a symmetric two-level fluctuator is $S_{sTLF} \propto [f_c^2+(2\pi f)^2]^{-1}$, where $f_c$ defines the corner frequency separating a low frequency $f^0$ dependence and a high frequency $f^{-2}$ dependence and characterizes the fluctuator time-scale \cite{schriefl2006}.  The power-law fit to the spectrum for device DA omits frequencies less than $2\times10^{-5}~\mathrm{Hz}$ (equivalent to time-scales of $1/f_c=13~\mathrm{hrs}$ which is consistent with Fig.~\ref{fig:drift}(b)) to characterize the noise roll-off.  We find $\beta=2.65$, which is greater than the expected value of 2.  This may be an artifact of the FFT as the time domain data indicates that the nature of the fluctuator is changing over the course of the measurement (fluctuator size is increasing with time).  The spectral density at $10^{-3}~\mathrm{Hz}$ for device DA is approaching that of devices DC, DE, and DF, so one may expect a return to $1/f$ behavior for slightly higher frequencies than explored here.

The drift characteristics of device DD does not fit the same characteristics as the rest of the devices.  It exhibits a much larger amplitude of drift, and the drift is not monotonic as in the case of the transient decay.  Nominally, device DD is similar to device DC in materials and the measurement temperature, with the differences being the device design and the quantity of donor implants.  We rule out the influence of the differing device design, as the noise characteristics in device DE (same design  and similar processing as device DD) are consistent with device DC.  The other differing aspect is that device DD received a donor implant density twice that of the other devices, which, coupled with a larger implant window, resulted in more than 3 times the donors being implanted.  While the other devices suggested that the implant process does not significantly effect the noise characteristics, there may be a threshold in the implant density or quantity for which the noise does become detrimentally effected.  It is also possible that the measurement system for this device introduced extra noise.

In summary, we have measured the low-frequency charge offset drift characteristics of intentionally implanted Si-MOS QD devices.  The devices have various lithographic designs and implant parameters.  In addition to equilibrium noise features, non-equilibrium features in the form of transient relaxation on the time scale of a few days and also discrete charge reconfigurations are present.  However, these non-equilibrium features are not dependent on the donor implants.  We note that the non-equilibrium features were not observed in previously measured SOI based Si-MOS QD devices.  The noise spectra indicate $1/f$ noise in the low-frequency range, as expected in Si-MOS devices, and devices with implanted donors exhibit noise magnitudes similar to best reported values in unimplanted Si-MOS, SiGe, and GaAs.  While there may be a detrimental effect on noise for high implant densities, modest implant densities provide a low-noise QD system for which a coupled QD-donor qubit can be accessed.

The authors wish to acknowledge M. D. Stewart Jr. and Binhui Hu from NIST for useful discussion during the preparation of this manuscript.  This work was performed, in part, at the Center for Integrated Nanotechnologies, an Office of Science User Facility operated for the U.S. Department of Energy (DOE) Office of Science. Sandia National Laboratories is a multi-mission laboratory managed and operated by National Technology and Engineering Solutions of Sandia, LLC., a wholly owned subsidiary of Honeywell International, Inc., for the U.S. Department of Energy's National Nuclear Security Administration under contract DE-NA-0003525.

\appendix
\section{Device Fabrication}

Phase 1 (silicon foundry): The initial material stack is fabricated using a $0.35~\mathrm{\mu m}$ silicon foundry process at Sandia National Laboratories. The starting material is a 150 mm diameter float zone $\langle 100 \rangle$ n-type silicon wafer with a room temperature resistivity of $>10~\mathrm{k\Omega cm}$. The two enriched silicon devices (DC and DD) start with a p-type float zone substrate with a $0.7~\mathrm{\mu m}$ thick epitaxial $\mathrm{^{28}Si}$ (500ppm $\mathrm{^{29}Si}$) layer. A $35~\mathrm{nm}$ thermal silicon oxide is grown at $900\mathrm{^\circ C}$ with dichloroethene (DCE) followed by a 30 min, $900\mathrm{^\circ C}$ $\mathrm{N_2}$ anneal. The next layer deposited is a $200~\mathrm{nm}$ amorphous silicon layer followed by a $5\times10^{15}~\mathrm{cm^{-2}}$, $35~\mathrm{keV}$ arsenic implant at $0^\circ$ tilt. The amorphous layers are crystallized later in the process flow to form a degenerately doped poly-silicon electrode. In the silicon foundry, the poly-Si is patterned and etched into a large scale region, a “construction zone” around $100\times100~\mathrm{\mu m^2}$ in size, that will later be patterned using e-beam lithography to form the nanostructure. After etching, ohmic implants are formed using optical lithography and implantation of As at $3\times10^{15}~\mathrm{cm^{-2}}$ density at $100~\mathrm{keV}$. An oxidation anneal of $900\mathrm{^\circ C}$ for 13 min and an $\mathrm{N_2}$ soak at $900\mathrm{^\circ C}$ for 30 min follow the implant step and serves the multiple purposes of crystallizing, activating and uniformly diffusing the dopants in the poly-Si while also forming a $\mathrm{SiO_2}$ layer (10-25 nm) on the surface of the poly-Si. This $\mathrm{SiO_2}$ layer forms the first part of the hard mask layer used for the nanostructure etch in the construction zone. The second part of the hard mask is a $35~\mathrm{nm}$ $\mathrm{Si_3N_4}$ layer. An $800~\mathrm{nm}$ thick field oxide is subsequently deposited using low pressure chemical vapor deposition (CVD) with tetraethoxysilane (TEOS) (this step is done by high density plasma CVD for the $\mathrm{^{28}Si}$ devices). The field oxide is planarized using chemical mechanical polishing (CMP) leaving approximately $500~\mathrm{nm}$ over the silicon and $300~\mathrm{nm}$ over the poly-Si. Vias are etched to the conducting poly-Si and n+ ohmics at the silicon surface. The vias are filled with Ti/TiN/W/TiN. The tungsten is a high contrast alignment marker for subsequent e-beam lithography steps. Large, approximately $100\times100~\mathrm{\mu m^2}$ windows aligned to the construction zones are then etched in the field oxide to expose the underlying hardmask and poly-Si construction zone for nanostructure patterning. The last processing step for the devices in the silicon foundry is a $450\mathrm{^\circ C}$ forming gas anneal for 90 min.
	
Phase 2 (nano-micro fabrication facility): The wafers are removed from the silicon foundry and subsequently diced into smaller parts, leading to $10~\mathrm{mm}\times 11~\mathrm{mm}$ die, each containing 4 complete QD devices. The nanostructures are patterned using electron beam lithography and a thinned ZEP resist. The pattern is transferred with a two-step etch process. First, the SiN and $\mathrm{SiO_2}$ hard mask layers are etched with a $\mathrm{CF_4}$ dry etch, and an $\mathrm{O_2}$ clean then strips the resist in-situ. The second etch step is to form the poly-Si electrodes, which is done with an HBr dry etch in the same chamber. The poly-Si etch is monitored using end-point detection in a large scale etch feature away from the active regions of the device. Wet acetone and dry $\mathrm{O_2}$ cleans are used to strip the residual resist after the poly-silicon nanostructure formation. After the wet strips of the tungsten vias, a lift-off process is used for aluminum formation of bond pads to contact the ohmics and poly-silicon electrodes. The last step is a $400\mathrm{^\circ C}$, 30 minute forming gas anneal. For devices that are implanted with donors near the QD region, a second e-beam lithography and implant step was done.  After the implant step (parameters provided in the main text), the photoresist was stripped with acetone and then the metal and residual organics were stripped from the surface using peroxide and RCA cleans, and then a dopant activation anneal was performed with the parameters indicated in the main text. The device was subsequently metallized using an Al lift-off process. 

Final material stack in the QD region is Si / $35~\mathrm{nm}$ $\mathrm{SiO_2}$ / $200~\mathrm{nm}$ poly-Si / $10-25~\mathrm{nm}$ $\mathrm{SiO_2}$ / $35~\mathrm{nm}$ $\mathrm{Si_3N_4}$.

\bibliographystyle{apsrev}
\bibliography{bib}

\end{document}